# A coupled neutronics shock physics solver: implementation of an SN neutron transport module

Eric Cervi, Antonio Cammi

*Politecnico di Milano, Via La Masa 34, 20156 Milan, Italy*
*eric.cervi@polimi.it, antonio.cammi@polimi.it*

**INTRODUCTION**

A multiphysics finite volume method (FVM) solver, coupling neutronics and shock physics, is under development at Politecnico di Milano for the analysis of shock imploding fissile materials [1]. The proposed solver can be a useful tool to make preliminary safety assessment of subcritical plutonium experiments [2] and, more in general, to perform criticality safety evaluations in case of strongly energetic events (such as chemical explosions) involving fissile materials [3].

To this aim, a multi-group $SP_3$ neutron transport model is coupled with a hydrodynamic shock physics model [4], suitable to describe the propagation of strong shockwaves in solid materials. The shock physics module implements a dynamic mesh to reproduce material deformations and its governing equations are written in an Arbitrary Lagrangian Eulerian (ALE) formulation to preserve the mesh quality in case of large distortions.

Different shock physics codes are available in literature, (see, e.g., [5,6]) but none of them implements a neutron transport module. Codes are also available to study non-linear wave propagations in liquid fuel reactors, but they are not suitable for shock compression of solids [7]. In this regard, the present solver is the only one coupling neutronics and shock physics models in the same simulation environment, without requiring external interfaces between different codes.

The purpose of this report is to improve the neutronics module of the coupled solver, by developing a discrete ordinate ($S_N$) model for the solution of the neutron transport equation. While the previously implemented $SP_3$ model has been successfully tested and verified in [8,9], the implementation of a more accurate neutronics model can improve the simulation of small imploding systems (as in the case of subcritical plutonium experiments), allowing for a better description of neutron leakages and of the flux behavior near the shock front, where density abruptly changes.

In the following sections, the new $S_N$ solver will be presented and tested on both steady state as well as transient case studies.

**Modelling approach**

In this section, the structure of the solver and the multi-physics coupling strategy are described. At each time step, the systems neutronics and shock physics are solved in two different iterative cycles, as shown in Figure 1. The temperature and density calculated by the shock physics module are passed to the neutronics one in order to evaluate cross sections. In turn, the fission power calculated by the neutronics solver appears as a source term in the energy equation in the shock physics module. External iterations between the neutronics and shock physics cycles are performed, to solve the non-linearities between the two physics. The neutronics and the shock physics models are briefly described in the following subsections. For more details, the reader is referred to [1].

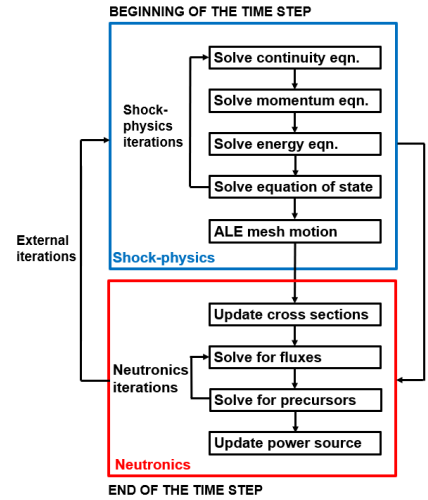

Fig. 1. Solver structure and coupling strategy.

*The shock physics model*

The implemented shock physics model is based on the *"hydrodynamic approximation"*. For very high pressures (above 5-10 GPa), the shear stresses become negligible and the solid response to shock compression is similar to that of an inviscid, compressible fluid [4]. Thanks to this approximation, the stress tensor and thermal conduction can be neglected in the conservation equations, which read as follows:

$$\frac{\partial \rho}{\partial t} + \nabla \cdot [\rho(\boldsymbol{u} - \boldsymbol{w})] = 0 \quad (1)$$

$$\frac{\partial (\rho \boldsymbol{u})}{\partial t} + \nabla \cdot [\rho \boldsymbol{u}(\boldsymbol{u} - \boldsymbol{w})] = -\nabla p + \boldsymbol{b} \quad (2)$$

$$\frac{\partial (\rho h)}{\partial t} + \nabla \cdot [\rho h(\boldsymbol{u} - \boldsymbol{w})] = \frac{Dp}{Dt} + \dot{q}_{fission} \quad (3)$$

where $\dot{q}_{fission}$ is the fission power heating, while the body force (gravity) is not considered for simplicity.

The balance equations are written in an Arbitrary Lagrangian-Eulerian (ALE) form: the mesh vertices can be moved with an arbitrary velocity $\boldsymbol{w}$, to preserve the mesh quality in case of strong distortions. This velocity is included in the advective terms of the equations in order to preserve the balances.

Under the hydrodynamic approximation, the material behavior can be described by the Mie-Gruneisen equation of state [4]:

$$p - p_H = \frac{\gamma(v)}{v}(e - e_H) \quad (4)$$

where $\gamma$ is the Gruneisen parameter, while $p_H$ and $e_H$ are the pressure and internal energy lying on a Hugoniot curve [4], which depends on the specific material and must be known experimentally.

For more details on the shock physics model the reader is referred to [1]. In this summary, two validation cases are reported, showing the agreement between the shock speed calculated by the model and the shock speed predicted by experimental Hugoniot curves for uranium and plutonium (Fig. 2).

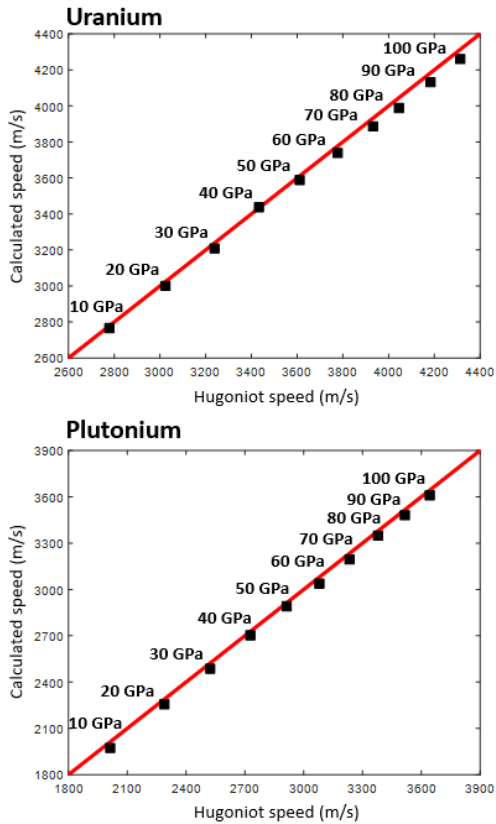

Fig. 2. Calculated (ordinate axis) vs. experimental (abscissa axis) shock speed for uranium and plutonium. The black squares represent plane shockwaves of different pressure. Perfect agreement is achieved if the squares lay on the red diagonal.

*The neutronics model*

In this section, the implemented $S_N$ neutronics model is presented. For a given neutron energy group $ei$ and a free-flight direction $di$, the neutron transport equation can be formulated as follows:

$$\frac{1}{v_{ei}}\frac{\partial \varphi_{ei,di}}{\partial t} + \nabla \cdot [(v_{ei}\boldsymbol{\Omega}_{di} - \boldsymbol{w}) \cdot n_{ei,di}] + \Sigma_{tot,ei}\varphi_{ei,di} = S_{f,ei,di} + S_{s,ei,di} + S_{d,ei,di} + Q_{ei,di} \quad (5)$$

where the arbitrary mesh velocity $\boldsymbol{w}$ appears in the divergence term, due to the ALE formulation of the solver. The effect of the solid particle motion on neutron flux [10] is not considered for simplicity. The finite volume method is used for the spatial discretization of the $S_N$ equations.

Both the number of energy groups and of flight directions can be arbitrarily selected by the user. The terms $S_{f,ei,di}$, $S_{s,ei,di}$ and $S_{d,ei,di}$ represent the fission source, the scattering neutrons and the delayed neutron source, respectively, and they are defined as follows:

$$S_{f,ei,di} = \sum_{ej,dj} \varphi_{ej,dj} w_{dj} [\nu_{ej}\Sigma_{f,ej}\chi_{p,ej}(1 - \beta_{tot})] \quad (6)$$

$$S_{s,ei,di} = \sum_{ej,dj,l} P_l(\boldsymbol{\Omega}_{di} \cdot \boldsymbol{\Omega}_{dj})(2l+1)\Sigma_{sl,ej \to ei}\varphi_{ej,dj}w_{dj} \quad (7)$$

$$S_{d,ei,di} = \sum_k \lambda_k c_k w_{di} \quad (8)$$

where $w_{di}$ is the weight of the direction $di$.

A source $Q_{ei,di}$ can also be considered for each energy group and direction combination. Its intensity and position can be defined by the user. If the neutron source is placed outside the computational domain, it can be accounted for as a boundary condition. This can be of interest for subcritical experiments in which the fissile sample is irradiated by an external source [11].

Transport equations are also included for precursors densities:

$$\frac{\partial c_k}{\partial t} + \nabla \cdot [c_k(\boldsymbol{u} - \boldsymbol{w})] = \beta_k \sum_i \bar{\nu}\Sigma_{f,i}\varphi_i - \lambda_k c_k \quad (9)$$

and a power iteration routine is implemented for the estimation of the multiplication factor.

**RESULTS**

**Steady-state verification**

In this section, 95% enriched uranium cubes of different dimensions are adopted as case studies. The multiplication factor of these cubes is evaluated using both the $SP_3$ and the $S_N$ modules and the results are compared to continuous energy Monte Carlo simulation. An $S_6$ angular discretization (i.e., 48 free-flight directions) and four energy groups (with cutoffs at 1, 2 and 3 MeV) are adopted, considering anisotropic scattering up to the seventh order. The directions and weights adopted in this work are based on the level symmetric quadrature sets given in [12]. Four energy groups are also selected for the SP$_3$ solver. On the other hand, Monte Carlo simulations are carried out using 100 million active neutron histories (10,000 cycles of 10,000 particles, plus 1000 inactive cycles to ensure fission source convergence).

The aim of this verification is to assess the capability of the two neutronics models to correctly predict reactivity in small systems, where neutron leakages are dominant and simpler models such as the SP$_3$ one may incur in significant limitations. Results are listed in Tables I and II.

TABLE I. S$_N$ vs. Monte Carlo results.

| Cube edge (cm) | $k_{eff}$ S$_N$ | $k_{eff}$ MC | Error (pcm) |
|---|---|---|---|
| 16 | 1.07376 | 1.07424 ± 0.00009 | -48 |
| 8 | 0.58075 | 0.57992 ± 0.00007 | +83 |
| 4 | 0.29227 | 0.29193 ± 0.00004 | +34 |
| 2 | 0.14519 | 0.14519 ± 0.00003 | 0 |
| 1 | 0.07222 | 0.07226 ± 0.00002 | -4 |

TABLE II. SP$_3$ vs. Monte Carlo results.

| Cube edge (cm) | $k_{eff}$ SP$_3$ | $k_{eff}$ MC | Error (pcm) |
|---|---|---|---|
| 16 | 1.06767 | 1.07424 ± 0.00009 | -657 |
| 8 | 0.58567 | 0.57992 ± 0.00007 | +575 |
| 4 | 0.29362 | 0.29193 ± 0.00004 | +169 |
| 2 | 0.14352 | 0.14519 ± 0.00003 | -167 |
| 1 | 0.07004 | 0.07226 ± 0.00002 | -222 |

Even using a relatively low number of flight directions, the S$_N$ model performs significantly better than the SP$_3$ one, always reducing the error with respect to Monte Carlo simulation well beyond 100 pcm.

**Shock implosion transient**

For demonstration purposes, the shock implosion of a plutonium sample is simulated, in order to highlight eventual differences between the S$_N$ and the SP$_3$ results. In more details, a 2D $^{239}$Pu cylinder with 3 cm radius is selected as a case study.

As a first step, a power iteration cycle is performed to determine the multiplication factor of the uncompressed cylinder. Again, a four-group S$_6$ model and a four-group SP$_3$ model are selected. While the former predicts an initial $k_{eff} = 0.90594$, the letter predicts $k_{eff} = 0.93651$. For comparison, a Monte Carlo simulation carried out with 100 million neutron histories yields a multiplication factor $k_{eff} = 0.90598 \pm 0.00008$, in excellent agreement with the S$_N$ result.

The fission source calculated by the power iteration cycle is, then, used as a constant source in the time dependent simulation, in order to have a non-zero neutron population, even if the cylinder is initially subcritical. To approximate zero power conditions, the fluxes evaluated by the power iteration cycle are normalized so that the average volumetric fission power is 1 W m$^{-3}$. In any case, it is reminded that different neutron sources can be arbitrarily defined by the user, depending on the specific application.

Once the initial condition at $t = 0$ is defined by means of the above procedure, a pressure of 30 GPa is applied to the external surface of the cylinder. The time evolution of the pressure field and of the fission power calculated by the S$_N$ and by the SP$_3$ models, adopting 50 picoseconds timesteps, are shown in Figs. 3, 4 and 5, respectively.

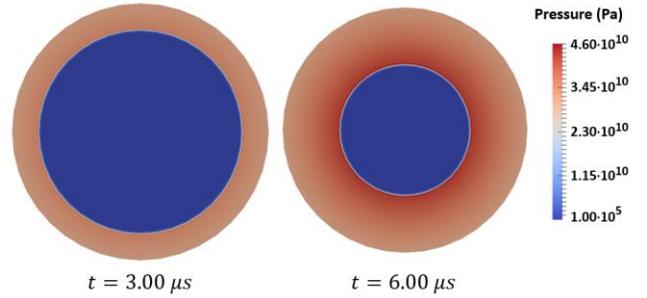

Fig. 3. Pressure field in the cylinder at $t = 3\ \mu s$ and $t = 6\ \mu s$.

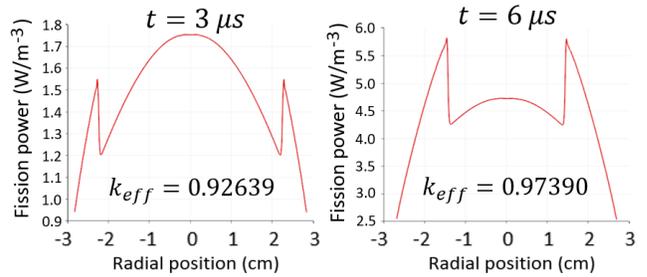

Fig. 4. Power density $t = 3\ \mu s$ and $t = 6\ \mu s$ (S$_N$ model).

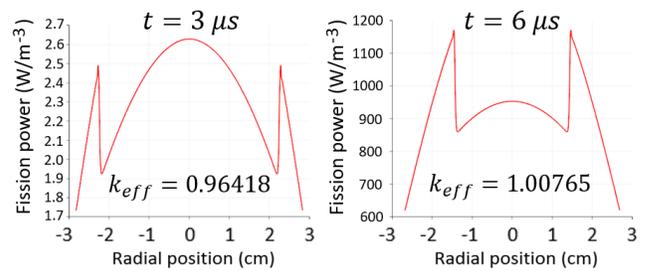

Fig. 5. Power density $t = 3\ \mu s$ and $t = 6\ \mu s$ (SP$_3$ model).

Different situations are predicted by the two neutronics models. According to the S$_N$ model, the cylinder is still subcritical at $t = 6\ \mu s$ (note that the fission power increases due to the presence of the source, even if $k_{eff} < 1$). On the other hand, according to the SP$_3$ model, critical mass is reached between $t = 3\ \mu s$ and $t = 6\ \mu s$.

The greater accuracy of the S$_N$ model in predicting the initial reactivity strongly benefits the evaluation of flux and reactivity evolution during the shock compression, leading to substantial differences compared to the simpler SP$_3$ approach. Due to the larger number of equations and implicit terms, the S6 runtimes increase by a factor 50, compared to the SP3 model.

## Conclusion

In the present work, a coupled multiphysics solver for the simulation of subcritical plutonium experiments and for criticality safety evaluation is extended by implementing a discrete ordinate neutronics model. The new model is significantly more accurate, compared to the previous one (multi-group SP$_3$ approximation of the neutron transport equation), achieving much smaller errors with respect to Monte Carlo simulations. Important differences are also highlighted in transient simulations, both in terms of reactivity as well as of fission power density.

The present solver can be a useful tool to design subcritical plutonium experiments and to predict whether critical mass is achieved during the shock implosion. In addition, for a given neutron source, it has the capability to estimate the energy release during the transient.

A possible future development can be the implementation of new material constitutive relations, in order to describe phenomena - such as spall fracture - that are not caught by the proposed hydrodynamic model. Another interesting extension could be the development of a chemical reaction module, in order to describe in deeper detail the chemical explosion leading to the shock compression.

## NOMENCLATURE

**Latin symbols**

| | |
|---|---|
| $\boldsymbol{b}$ | Body force, kg m$^{-2}$ s$^{-2}$ |
| $c$ | Precursor density, m$^{-3}$ |
| $h$ | Enthalpy, J kg$^{-1}$ |
| $n$ | Neutron density, m$^{-3}$ |
| $P_l$ | Legendre polynomial ($l^{th}$ order) |
| $p$ | Pressure, Pa |
| $Q$ | Neutron source, m$^{-3}$ s$^{-1}$ |
| $\dot{q}$ | Power source, J s$^{-1}$ m$^{-1}$ |
| $t$ | Time, s |
| $\boldsymbol{u}$ | Material velocity, m s$^{-1}$ |
| $v$ | Specific volume, m$^3$ kg$^{-1}$ |
| $v_i$ | Neutron velocity, m s$^{-1}$ |
| $\boldsymbol{w}$ | Arbitrary mesh velocity, m s$^{-1}$ |

**Greek symbols**

| | |
|---|---|
| $\beta$ | Delayed neutron fraction, - |
| $\gamma$ | Grüneisen parameter, - |
| $\lambda$ | Precursor decay constant, s$^{-1}$ |
| $\bar{\nu}$ | Average neutrons per fission, - |
| $\rho$ | Density, kg m$^{-3}$ |
| $\Sigma$ | Macroscopic cross section m$^{-1}$ |
| $\chi$ | Neutron yield, - |
| $\boldsymbol{\Omega}$ | Flight direction, - |

**Subscripts**

| | |
|---|---|
| $d$ | Delayed |
| $ei$ | Neutron energy group index |
| $di$ | Flight direction index |
| $f$ | Fission |
| $H$ | Hugoniot |
| $k$ | Delayed neutron precursor group |
| $p$ | Prompt |
| $r$ | Removal |
| $sl$ | Inelastic scattering ($l^{th}$ order) |
| $t$ | Total |
| $tr$ | Transport |